\documentclass[12pt]{article}
\pdfoutput=1

\usepackage{graphicx}
\usepackage{dsfont}
\usepackage{amssymb}
\usepackage{mathrsfs}

\begin{document}
\title{Superantenna made of transformation media}
\author{Ulf Leonhardt$^1$ and Tom\'a\v{s} Tyc$^{1,2}$\\
$^1$School of Physics and Astronomy, University of St Andrews,\\
North Haugh, St Andrews KY16 9SS, Scotland\\
$^2$Institute of Theoretical Physics and Astrophysics,\\ 
Masaryk University, 
Kotlarska 2, 61137 Brno, Czech Republic
}
\date{\today}
\maketitle
\begin{abstract}
We show how transformation media can make a superantenna that
is either completely invisible or focuses incoming light into 
a needle-sharp beam. Our idea is based on representating
three-dimensional space as a foliage of sheets and performing
two-dimensional conformal maps on each sheet.

\end{abstract}

\newpage

\section{Introduction}

Transformation optics 
\cite{Greenleaf1,Greenleaf2,LeoConform,LeoNotes,PSS,GREE,Review} 
has the potential to transform optics, both 
by developing new ideas for the design of optical instruments
and by stimulating research into new optical materials.
Transformation optics uses the fact \cite{GREE}
that optical media alter the geometry of space and time for 
light.\footnote{For a review see Ref.\  \cite{Review}.}
A transformation medium performs an active coordinate transformation:
electromagnetism in physical space, 
including the effect of the medium, 
is equivalent to electromagnetism in transformed coordinates 
where space appears to be empty. 
The sole function of the device is to facilitate a transformation
from physical to electromagnetic space.
One of the most striking applications of transformation optics
is invisibility 
\cite{Greenleaf1,Greenleaf2,LeoConform,PSS};
a cloaking device creates a coordinate transformation 
that carries a hole for hiding objects.
But invisibility is by no means the only application:
perfect lensing based on negative refraction also implicitly uses
coordinate transformations \cite{GREE}
 --- multivalued transformations --- 
although the original argument \cite{Pendrylens}
why negative refraction makes a perfect lens
did not make use of this fact. 
Similarly, the performance of cloaking devices can be 
theoretically verified
by other means than coordinate transformations 
\cite{Cloakingscattering},
but the ideas of transformation optics provide an elegant,
visual way of designing optical devices.
Such visual thinking often leads to new insights
and inventions.

Here we discuss an idea for a superantenna based on 
coordinate transformations.
This device is either invisible or it condenses 
a cross section of light into a single point.
Ideally, the condensed light would continue 
to propagate as a needle-sharp beam in a given direction,
but there might be practical limitations to this behavior.
Our idea certainly is not the only case where optical condensors
based on transformation media have been described
\cite{LeoConform,LeoNotes,Condensors},
but it may have unique three-dimensional properties,
because it changes the topology of space.
Most nontrivial applications of transformation media 
use topology transformations: 
spaces with holes
in the case of invisibility \cite{GREE}, 
multivalued spaces for 
perfect lensing \cite{GREE},
topological bridges for electromagnetic wormholes
\cite{Wormholes}
and space-time horizons \cite{GREE}
for artificial black holes.   
Our idea relies on a three-dimensional extension 
of optical conformal mapping \cite{LeoConform,LeoNotes}
as we describe below.

\section{Transformation}

Imagine we represent three-dimensional space as a foliage
of spherical layers (such that it resembles an onion), 
as illustrated in Fig.\ \ref{fig:onion}.
Each layer we map onto an individual plane 
by stereographic projection 
\cite{LeoNotes,Needham,Ablowitz},
each layer is a Riemann sphere \cite{Needham,Ablowitz},
see Fig.\ \ref{fig:stereo}.
Expressed in terms of analytical geometry, 
the Cartesian coordinates $X$, $Y$, $Z$ of physical space
are represented by spherical radii $r$ and 
dimensionless complex numbers $z$ that correspond to
the stereographic projections of the spherical angles.
In formulae,
%%%%%%
\begin{equation}
r = \sqrt{X^2+Y^2+Z^2}
\,,\quad
z = \frac{X+iY}{r-Z} \,.
\label{eq:stereo}
\end{equation}
%%%%%%
Then we perform a conformal transformation on $z$,
mapping $z$ to $w$ where the mapping may vary with the 
radius $r$, but $w$ should only depend on $z$ and not on $z^*$,
%%%%%%
\begin{equation}
w = w(z,r) \,,\quad
\frac{\partial w}{\partial z^*} = 0 \,.
\label{eq:wmap}
\end{equation}
%%%%%% 
The main advantage of such restricted maps is that they 
allow us to use the plethora of tools and insights of 
complex analysis, because the condition 
${\partial w}/{\partial z^*} = 0$
corresponds to the Cauchy-Riemann differential equations
of analytic functions
\cite{Ablowitz}.
The map $w(z)$ generates a Riemann surface that consists of
several Riemann sheets, 
unless $w(z)$ is a M\"obius transformation 
\cite{Needham}.
Finally we map the $w$ sheets back on the Riemann sphere 
of electromagnetic space
by the inverse stereographic projection
\cite{Needham,Ablowitz}
%%%%%%
\begin{equation}
X'+iY' = \frac{2rw}{|w|^2+1}
\,,\quad
Z' = r\,\frac{|w|^2-1}{|w|^2+1} \,.
\label{eq:em}
\end{equation}
%%%%%%
On each individual Riemann sphere,
the transformation ({\ref{eq:wmap}) is conformal 
\cite{Needham,Ablowitz,Nehari}
--- it preserves the angles between lines ---
but the transformation is normally 
not conformal in three-dimensional space,
where the Cartesian coordinates $X$, $Y$, $Z$
are mapped onto new coordinates
$X'$, $Y'$, $Z'$ and vice versa.
In the spirit of transformation optics \cite{GREE}, 
the new coordinates are the coordinates of a virtual space,
electromagnetic space,
that we assume to be completely empty, where 
light propagates along straight lines.
The device transforms the straight light rays of electromagnetic space
into appropriately curved rays in physical space.
Moreover, it not only transforms rays,
but complete electromagnetic waves.  

An example of a sufficiently simple but interesting 
conformal transformation is the Joukowski map 
(Zhukovsky map) known from
aeronautical engineering \cite{Aero}
%%%%%%
\begin{equation}
w = z + \frac{a^2(r)}{z} 
\,,\quad
z = \frac{1}{2}\left(w\pm\sqrt{w^2-4a^2(r)}\right) \,.
\label{eq:w}
\end{equation}
%%%%%%
Equation (\ref{eq:w}) describes a conformal map
between the $z$ plane and two Riemann sheets in $w$ space.
Or, in terms of Riemann spheres, for each radius $r$, two spheres
in electromagnetic space are merged into one sphere
in physical space, as illustrated in Fig.\ \ref{fig:maps}.
The image of each one of the two Riemann spheres 
occupies a certain spherical
angle in physical space.

This topology makes a perfect trap: 
in electromagnetic space light may pass 
from one of the two spheres to the other, 
and it will stay there. 
In physical space, however, the spherical angle 
of the second Riemann sphere
may shrink to a point where light is concentrated.
The entrance to the trap is marked by the branch points in $w$ space,
{\it i.e.}\ by the points $\pm 2a(r)$.
These points are variable, because $a$ is a function of the radius $r$.
Therefore, one could catch incoming light by rapidly making $a(r)$
very large when the light gets sufficiently close,
say for $r$ smaller than a critical radius $r_0$.
When the light has passed from the first Riemann sphere to the second one
it will continue to propagate on the second sphere.
As the radius $r$ grows again, the fraction of real space 
occupied by this sphere is diminishing to zero,
the trap closes:
light is concentrated and funneled into a needle-sharp beam.

What is the cross section of the captured light? 
The cross section in physical space is identical to the cross section 
in electromagnetic space, because, outside of the device,
physical and electromagnetic space are identical. 
In electromagnetic space light rays travel along straight lines. 
Here the portal to the trap is marked by the line of the branch points $\pm2a(r)$.
Consequently, the cross section in real space is bounded by 
the curve of the branch points in electromagnetic space,
by the inverse stereographic projection of $\pm2a(r)$.
Figure\ \ref{fig:heart} shows an example that is given by 
the radius-dependent branch points 
%%%%%%
\begin{equation}
a = \left\{
\begin{array}{r@{\quad:\quad}l}
a_0\left( \frac{\displaystyle r-r_0}{\displaystyle r_0}\right)^2 & r \le r_0 \\
0 & r > r_0 \end{array}
\right.
\label{eq:a}
\end{equation}
%%%%%%
where $a_0$ describes the maximal extension of the branch cut
at $r=0$.
The captured light has entered the second Riemann sphere
in electromagnetic space that shrinks into the point $z=0$
on the Riemann sphere in physical space when $r$ approaches $r_0$. 
So, in physical space, the captured light rays are  
concentrated in the spatial direction
given by the image of $z=0$ 
in the inverse stereographic projection,
{\it i.e.}\ along the negative $Z$ axis. 
Figure \ref{fig:heart} shows how the device 
funnels a captured light ray towards this axis; 
Fig.\ \ref{fig:rays} illustrates the fate of a circle of light rays. 
The device acts like a superantenna, 
focusing incident electromagnetic radiation into an area
that is, in principle, infinitely small. 
Moreover, and quite remarkably, 
the focus is independent of the direction of incidence,
in contrast to the focusing by normal lenses or mirrors.
The theory would suggest that the captured electromagnetic radiation 
is concentrated in an infinitely narrow beam. 
In practice, of course, both the focus and the beam will be limited,
but assessing these limitations goes beyond the scope of this paper,
beyond the area of transformation optics. 

\section{Media}

What are the requirements to implement such a superantenna in practice?
Perhaps the most elegant way of calculating the material properties
of this device is using the connection between transformation media
and general relativity \cite{GREE,Review}.
For this, we need to establish the geometries that correspond
to the stereographic coordinates (\ref{eq:stereo})
and their transformation (\ref{eq:wmap}).
We denote the metric tensor of the stereographic coordinates by
$\Gamma$ and the metric tensor of the transformed coordinates by $G$
(using the notation of Ref.\ \cite{Review}). 
The line element in physical space is given by
%%%%%%
\begin{equation}
\mathrm{d}l^2 = 
\mathrm{d}X^2 + \mathrm{d}Y^2 + \mathrm{d}Z^2 =
\mathrm{d}r^2 + \frac{4r^2\mathrm{d}z\,\mathrm{d}z^*}
{(|z|^2+1)^2} \,.
\label{eq:line}
\end{equation}
%%%%%%
The geometry of the transformed coordinates is characterized 
by the equivalent line element (\ref{eq:line}) in electromagnetic space, 
but expressed in real-space coordinates, 
%%%%%%
\begin{equation}
\mathrm{d}s^2 = 
\mathrm{d}r^2 + \frac{4r^2\mathrm{d}w\,\mathrm{d}w^*}
{(|w|^2+1)^2} 
\quad\mbox{with}\quad
\mathrm{d}w = \frac{\partial w}{\partial r}\,\mathrm{d}r
+  \frac{\partial w}{\partial z}\,\mathrm{d}z
\,.
\label{eq:wline}
\end{equation}
%%%%%%
For formulating a compact notation for the metric in electromagnetic space
we define 
$\nabla \equiv (\partial/\partial r, \partial/\partial z, \partial/\partial z^*)$
such that, for example, 
$\nabla w = (\partial w/\partial r, \partial w/\partial z, 0)$.
From the line element (\ref{eq:wline})
we read off the matrix of the metric
%%%%%%
\begin{equation}
G = \nabla r \otimes \nabla r +
\frac{2r^2\left( \nabla w \otimes \nabla w^* 
+ \nabla w^* \otimes \nabla w \right)}{(|w|^2+1)^2}
\,.
\label{eq:gs}
\end{equation}
%%%%%%
We obtain the metric tensor $\Gamma$ of the stereographic coordinates
of real space by putting $w=z$ in $G$ or by directly reading off the 
tensor from the line element (\ref{eq:line}).  
The square root of the determinant of the metric tensor describes
the volume element.
We put the matrix (\ref{eq:gs}) in explicit form and calculate
the determinant:
%%%%%%
\begin{equation}
g=\mathrm{det} G =- \frac{4r^4}{(|w|^2+1)^4}
\left|\frac{\partial w}{\partial z}\right|^4
\,,\quad
\gamma=\mathrm{det} \Gamma = - \frac{4r^4}{(|z|^2+1)^4}
\,.
\label{eq:dets}
\end{equation}
%%%%%%
The inverse matrix $G^{-1}$ describes the contravariant
metric tensor and is given by
%%%%%%
\begin{equation}
G^{-1} = 
\frac{\partial \mathbf{z}}{\partial r} \otimes
\frac{\partial \mathbf{z}}{\partial r} +
\frac{(1+|w|^2)^2}{2r^2}
\left(
\frac{\partial \mathbf{z}}{\partial w} \otimes
\frac{\partial \mathbf{z}}{\partial w^*} +
\frac{\partial \mathbf{z}}{\partial w^*} \otimes
\frac{\partial \mathbf{z}}{\partial w}
\right)
\label{eq:invgs}
\end{equation}
%%%%%%
with $\mathbf{z}$ defined as the vector of $(r,z,z^*)$
and $z$ understood as a function of $w$ and $r$.
One can verify 
that Eq.\ (\ref{eq:invgs}) indeed
describes the inverse matrix of $G$
using the relations
%%%%%%
\begin{equation}
\nabla z^i \cdot \frac{\partial\mathbf{z}}{\partial z^j} = 
\delta_j^i 
\,,\quad
\sum_i \nabla z^i \otimes \frac{\partial \mathbf{z}}
{\partial z^i} = 
\nabla \otimes \mathbf{z} = \mathds{1}
\,.
\end{equation}
%%%%%%
According to the connection between optical
media and geometry \cite{GREE,Review}
the contravariant electric permittivity tensor in 
stereographic coordinates is
%%%%%%
\begin{equation}
\varepsilon  =
\frac{\sqrt{-g}} {\sqrt{-\gamma}} \,G^{-1} =
\frac{(|z|^2+1)^2}{(|w|^2+1)^2}
\left|\frac{\partial w}{\partial z}\right|^2 G^{-1}
\,.
\label{eq:eps}
\end{equation}
%%%%%%
The magnetic permeability tensor should be identical to $\varepsilon$
(impedance matching).  
The matrix (\ref{eq:eps}) describes a contravariant tensor;
the eigenvalues of $\varepsilon$ 
are not invariant under coordinate transformations,
but the eigenvalues of the mixed tensor $\varepsilon \Gamma$ are \cite{Review}.
Therefore they correspond to the eigenvalues of the 
electric permittivity in Cartesian coordinates
that characterize the dielectric response of the anisotropic 
material used to implement the coordinate transformation 
(\ref{eq:wmap}).
We calculate the three eigenvalues of  $\varepsilon \Gamma$ and obtain
%%%%%%
\begin{equation}
\varepsilon_1  = 1
\,,\quad 
\varepsilon_\pm = 
\frac{1}{2}\left(A\pm\sqrt{A^2-4B}\right)
\label{eq:eigen}
\end{equation}
%%%%%%
with the abbreviations
%%%%%%
\begin{equation}
A = 1 + B \left(1+2\sqrt{-\gamma}\,
\left|\frac{\partial w}{\partial r}\right|^2
\left|\frac{\partial w}{\partial z}\right|^{-2}
\right)
\,,\quad
B = \frac{\sqrt{-g}} {\sqrt{-\gamma}}
\,.
\end{equation}
%%%%%%
One of the eigenvalues, $\varepsilon_1$,
is always unity,
which reflects the fact that the coordinate transformations
(\ref{eq:wmap}) act on two-dimensional manifolds,
although they cover layer-by-layer the entire
three-dimensional physical space as shown in Fig.\ \ref{fig:onion}.
Close to the focal line of the device at $z=0$ (in negative $Z$ 
direction) we would expect that the eigenvalues 
of the dielectric tensor become singular.
We find for the transformation (\ref{eq:w})
%%%%%%
\begin{equation}
\varepsilon_+ \rightarrow \frac{1}{a^4}
\,,\quad
\varepsilon_- \rightarrow 1
\quad\mbox{for}\quad z  \rightarrow 0
\,.
\end{equation}
%%%%%%
Since $a(r)$ vanishes at $r_0$ one of the eigenvalues
indeed is singular.
In practice, the dielectric functions will always be finite,
which limits the achievable focus of the antenna 
and may also influence the width of the beam 
funneled into the negative $Z$ direction.
In order to estimate these effects and to develop the 
optimal strategy for regularizing the singularity at the focus,
one could perhaps use similar mathematical techniques
as the ones applied for assessing the regularization
of invisibility devices \cite{Lassas}.

\section{Outlook}

Our principal ideas are by no means restricted to superantennae,
but they could serve to design other novel optical devices as well.
One could imagine other foliages of three-dimensional space
than the Riemann spheres of Fig.\ \ref{fig:onion} 
and consider conformal maps on them.
The important point is that such three-dimensional foliages
allow researchers to make use of the tools of two-dimensional 
conformal mapping. 
In this way, a rich arsenal of insights, techniques and tools 
becomes available
for designing optical instruments,
the theory of complex analysis and conformal mapping 
\cite{Needham,Ablowitz,Nehari}
that has reached a high level of development
and sophistication, and combines, like no other branch 
of mathematics, analytical techniques with visual thinking.

We are grateful to T.\ G.\ Philbin for helpful comments. 
The paper has been supported by 
COVAQIAL and
a Royal Society Wolfson Research Merit Award.

%%%

\newpage
%%%
\begin{figure}[h]
\begin{center}
\includegraphics[width=17.0pc]{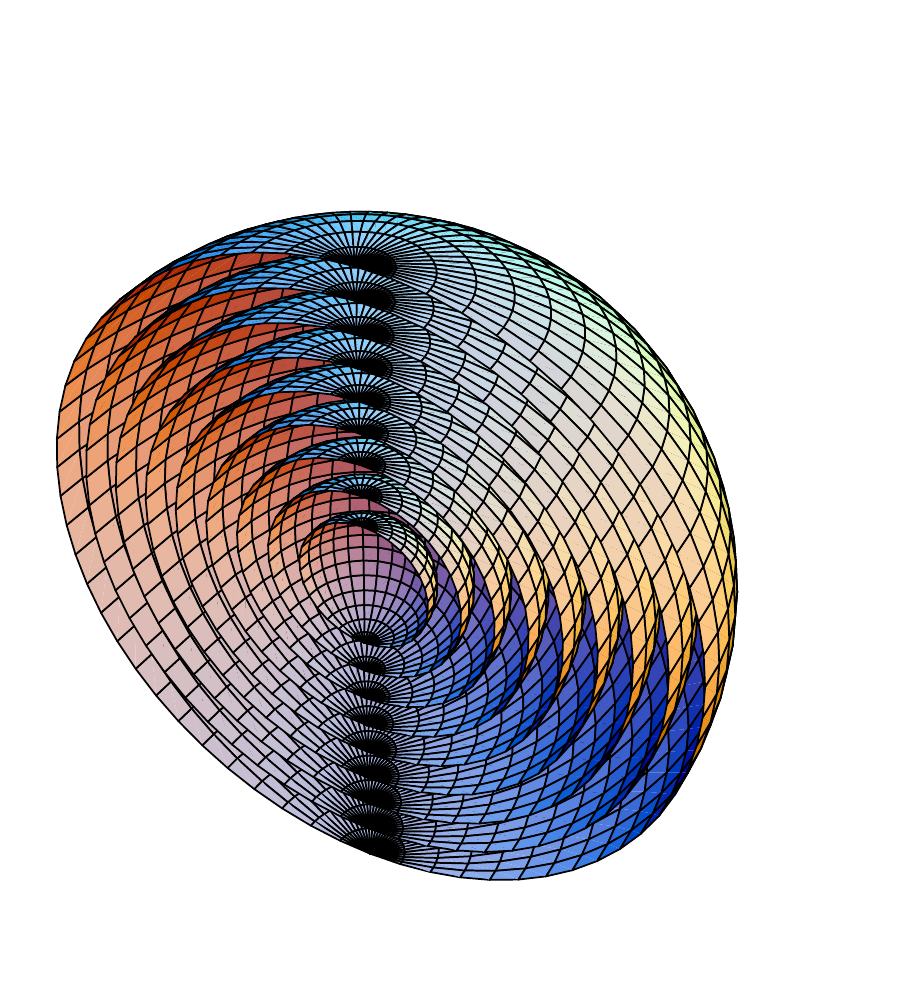}
\caption{{\small
Foliage. 
We present three-dimensional space as a foliage of spherical layers
(artificially cut in half to be seen in the picture).
Each layer represents a Riemann sphere.
A device that performs conformal mappings on these spheres may 
act like a superantenna.  
}
\label{fig:onion}
}
\end{center}
\end{figure}
%%%

\newpage

%%%
\begin{figure}[h]
\begin{center}
\includegraphics[width=18.0pc]{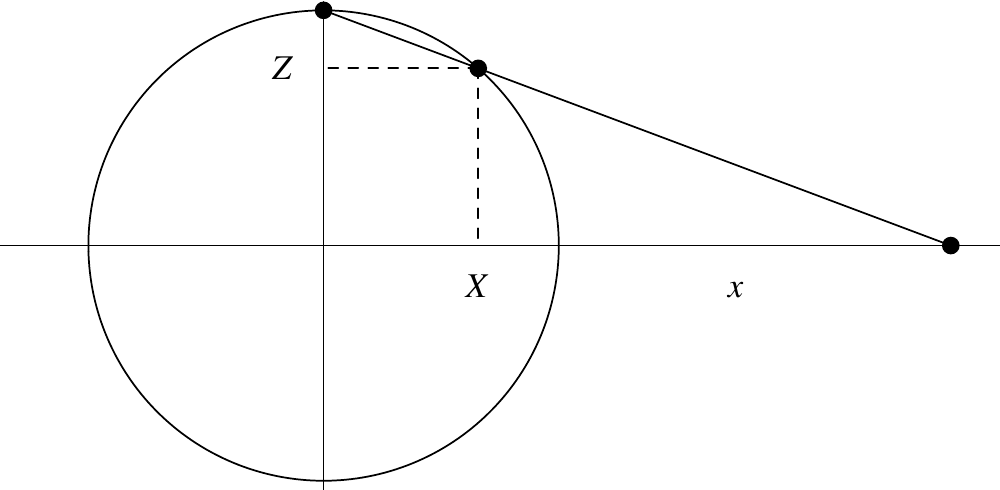}
\caption{{\small
Stereographic projection,
mapping the $(x,y)$
plane onto the  $(X,Y,Z)$ surface of a sphere,
shown here at $Y=0$ where the plane appears as a line
and the sphere as a circle.
A line drawn from the North Pole of the sphere 
through $(X,Y,Z)$
cuts the plane at $(x,y)$.
Conversely,
in the inverse stereographic projection, 
the line between the North Pole and $(x,y)$ 
cuts the surface of the sphere at $(X,Y,Z)$.
}
\label{fig:stereo}
}
\end{center}
\end{figure}
%%%

\newpage

%%%
\begin{figure}[h]
\begin{center}
\includegraphics[width=30.0pc]{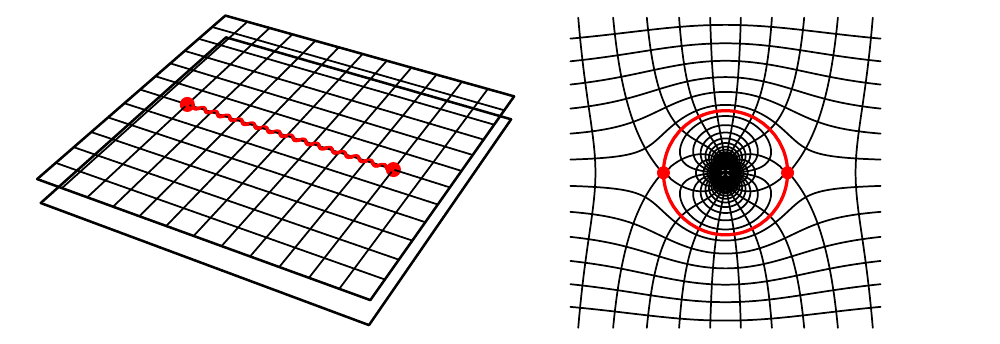}\\
\includegraphics[width=30.0pc]{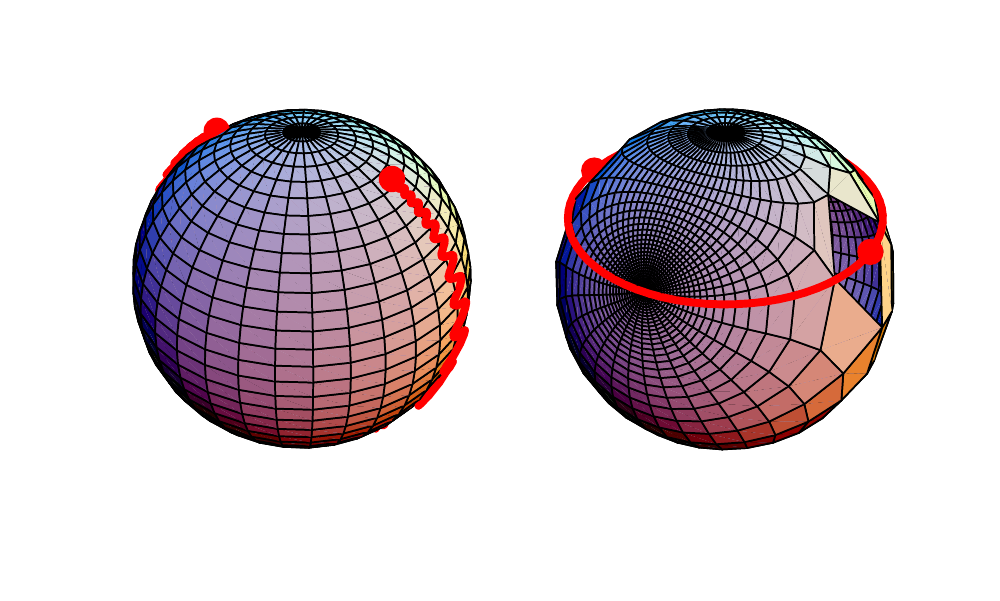}
\caption{{\small
Conformal map of the sphere.
The top pictures illustrate the conformal map (\ref{eq:w})
on the complex plane, while the bottom pictures show 
the corresponding map on the sphere.
The red points are the branch points that are connected 
by the branch cut, also shown in red.
The pictures on the left column 
illustrate the topology of electromagnetic space,
while the pictures on the right describe physical space.  
Light that passes from one branch of electromagnetic space
to another is trapped in the corresponding region in 
physical space. 
When light rays approach the device the branch cut opens;
it closes when the rays leave, but then they are trapped 
on another Riemann sheet that shrinks to zero in physical space:
light is concentrated in a given direction. 
The map makes a perfect trap.
}
\label{fig:maps}
}
\end{center}
\end{figure}
%%%

\newpage

%%%
\begin{figure}[h]
\begin{center}
\includegraphics[width=40.0pc]{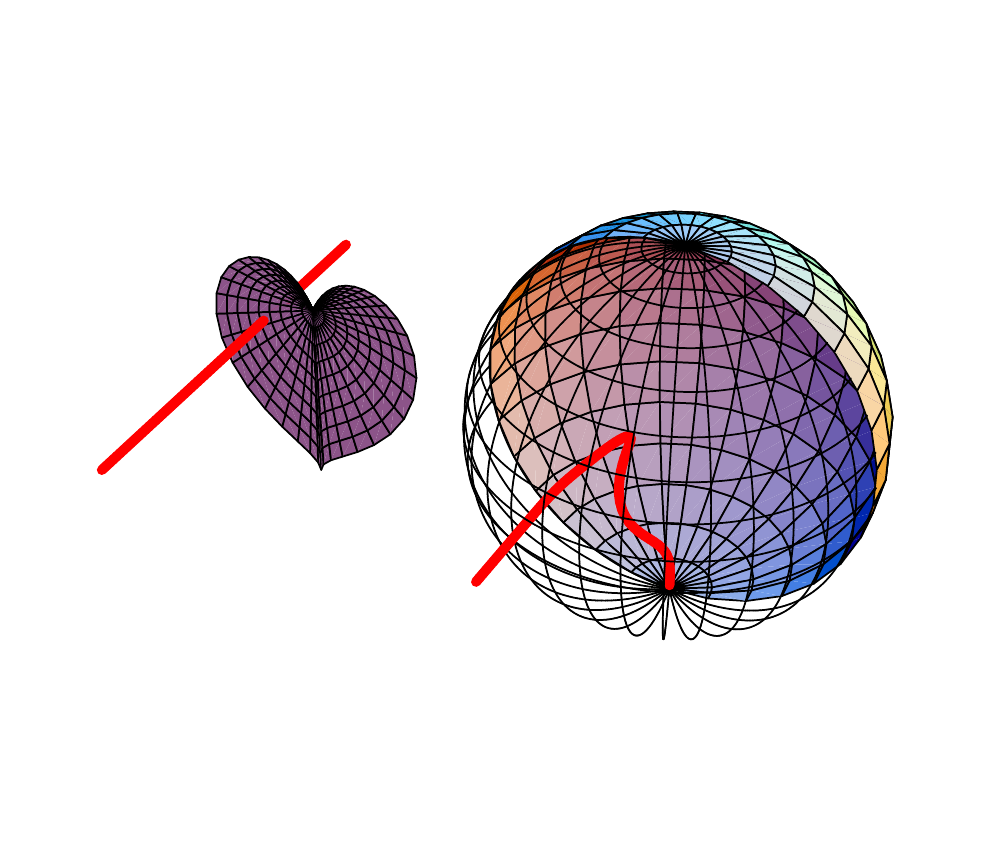}
\caption{{\small
Light capture.
The left picture shows a light ray in electromagnetic space, 
whereas the right picture displays the ray 
trajectory in real space. 
We used Eq.\ \ref{eq:a} with the parameters $r_0=5$ and $a_0=2.5$.
The heart-shaped leaf illustrates the cross section 
of light capture;
any ray that hits this area is captured.
The right figure shows how the light ray is bent and funneled 
towards the bottom of the device where
all captured light rays would converge
in a needle-sharp beam. 
}
\label{fig:heart}
}
\end{center}
\end{figure}
%%%

\newpage

%%%
\begin{figure}[h]
\begin{center}
\includegraphics[width=30.0pc]{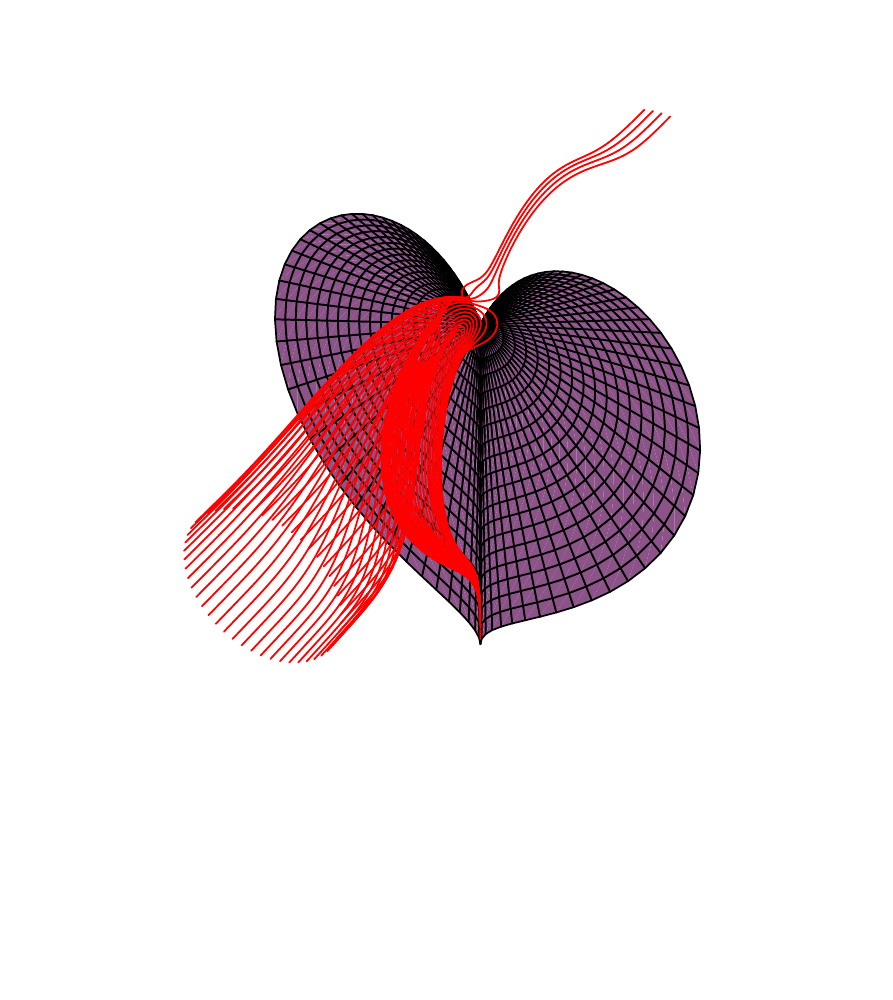}
\caption{{\small
Capture of light rays.
A circle of light rays enter the superantenna
(indicated by the heart-shaped capture cross section).
The rays within the cross section are directed 
towards the bottom of the device and bundled in an
extremely narrow beam. 
The rays that escape
continue in the original directions;
for them the device is invisible.
}
\label{fig:rays}
}
\end{center}
\end{figure}
%%%

\end{document}